# Synchronous multi-color laser network with daily sub-femtosecond timing drift

Kemal Şafak,[1,2] Ming Xin,[1,3,*] Michael Y. Peng,[3] and Franz X. Kärtner[1,3,4]

[1]*Center for Free-Electron Laser Science, Deutsches Elektronen-Synchrotron, Notkestrasse 85, Hamburg 22607, Germany*
[2]*Cycle GmbH, Notkestrasse 85, Hamburg 22607, Germany*
[3]*Research Laboratory of Electronics, Massachusetts Institute of Technology, Cambridge, Massachusetts 02139, USA*
[4]*Department of Physics and the Hamburg Center for Ultrafast Imaging, Luruper Chaussee 149, 22761 Hamburg, Germany*
*Corresponding author: xinm@mit.edu

Filming atoms in motion with sub-atomic spatiotemporal resolution is one of the distinguished scientific endeavors of our time. Newly emerging X-ray laser facilities are the most likely candidates to enable such a detailed gazing of atoms due to their angstrom-level radiation wavelength. To provide the necessary temporal resolution, numerous mode-locked lasers must be synchronized with ultra-high precision across kilometer-distances. Here, we demonstrate a metronome synchronizing a network of pulsed-lasers operating at different center wavelengths and different repetition rates over 4.7-km distance. The network achieves a record-low timing drift of 0.6 fs RMS measured with 2-Hz sampling over 40 h. Short-term stability measurements show an out-of-loop timing jitter of only 1.3 fs RMS integrated from 1 Hz to 1 MHz. To validate the network performance, we present a comprehensive noise analysis based on the feedback flow between the setup elements. Our analysis identifies nine uncorrelated noise sources, out of which the slave laser's inherent jitter dominates with 1.26 fs RMS. This suggests that the timing precision of the network is not limited by the synchronization technique, and so could be much further improved by developing lasers with lower inherent noise.

**Introduction**

It has been more than fifty years since the pioneers of laser science were able to lock the longitudinal modes in a He-Ne laser for the first time[1]. Over the following years, the discovery of passively mode-locked lasers[2] has laid the foundations of two thriving scientific fields: high-energy ultrashort optical pulse generation and optical frequency metrology. The former has thrived with the invention of chirped-pulse amplification[3] and broadband dispersion control with chirped mirrors[4,5]. This led to the emergence of new research fields such as attosecond physics[6] and laser-based particle acceleration[7]. Parallel to these achievements, optical frequency metrology has flourished with the invention of optical frequency combs[8] which have pushed the frequency measurement instabilities down to the 19th decimal place[9] and led to the development of optical clocks[10]. Furthermore, it is proven that a well-defined optical frequency comb with ultra-sharp lines owes its existence to the perfect temporal periodicity and amplitude stability of the pulse train generated by the mode-locked laser[11]. Today, researchers are able to compare the frequency of distant optical clocks located in different countries[12] which could soon lead to the redefinition of the second[13]. Five decades since their emergence, these two evolving scientific fields are now crossing paths once more to solve the challenging task of capturing ultrafast structural dynamics on the scale of atoms and electrons. The realization of this long-standing scientific aspiration requires two fundamental advances: i) ultra-short, high-flux, coherent hard

X-ray sources and ii) an ultra-precise, highly stable timing distribution to provide the required spatial and temporal resolution respectively.

Emerging laser-based attoscience facilities (e.g., the Extreme Light Infrastructure[14]) and X-ray free electron lasers (XFEL) (e.g., the European XFEL[15], LCLS II[16]) are already offering solutions to the first challenge of producing efficient hard X-rays at high repetition rates. However, they also require facility-wide tight synchronization of the driving lasers (e.g., injector-, seed- and pump-laser etc.) to generate ultra-short, stable X-ray pulses[17] and to enable high-precision time-resolved optical pump, X-ray probe experiments[18]. Therefore, the second challenge necessitates the development of a synchronous mode-locked laser network similar to the one described in Fig. 1. Here, the timing signal from a master laser is transferred via a timing-stabilized fiber link network to synchronize various remote slave lasers with daily sub-femtosecond relative timing drifts.

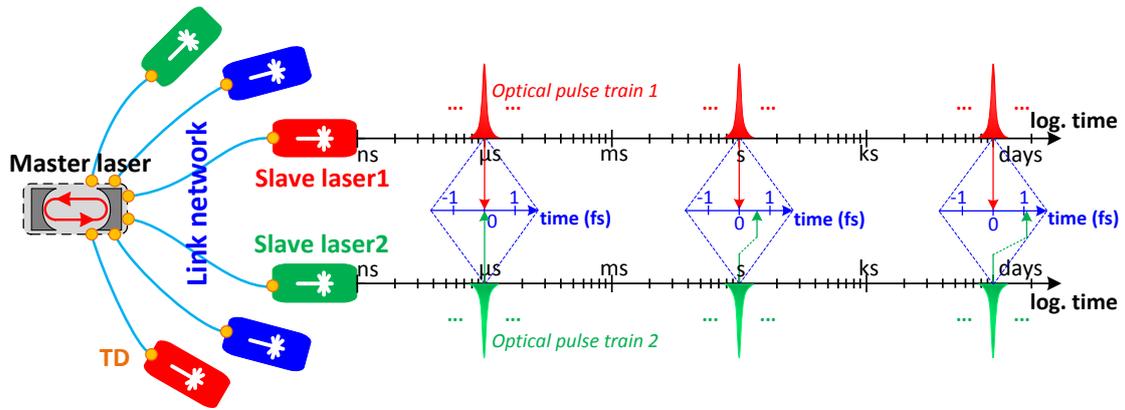

**Figure 1.** Layout of a synchronous mode-locked laser network with sub-femtosecond daily drift. Temporal duration and periodicity of the red and green pulses are not to scale with the time axes. The time axes under the pulses are in logarithmic scale, whereas the time axes in the middle are in linear scale on the order of femtoseconds. Orange filled circles symbolize timing detectors (TD).

To construct such a network, attosecond precision timing detectors[19,20], ultra-low noise mode-locked lasers[11,21], drift-free fiber network transmission[22,23], and tight synchronization of the remote slave lasers[24-26] are needed. Although the feasibility of these individual elements has already been shown, there have been no reports of a multi-color, long-term stable synchronous mode-locked laser network featuring these components. To achieve this, utmost attention must be given to every possible noise source in the system. For example, timing errors may still exist at the output of the laser network; even if perfect synchronization is achieved, i.e., the timing signal of the master laser is transmitted to the slave lasers without any noise added by the system. If not accounted for, slow drifts in the master laser's repetition rate can cause femtosecond-level timing drifts. The most stringent requirement on the long-term stability of the master laser's repetition rate stems from the "uncompensated" beam paths, such as short reference arms used for time-of-flight stabilization whose lengths cannot be varied. In this case, Eq. (1) gives the out-of-loop timing error due to the deviation of the master laser's repetition rate (derivation given in the Supplementary Information):

$$\Delta t = \frac{\Delta f_{rep}}{f_{rep}} \frac{\Delta d}{c} \qquad (1)$$

where $f_{rep}$ is the repetition rate of the master laser, $\Delta f_{rep}$ is its deviation, $\Delta d$ is the length difference of the uncompensated beam paths between two fiber links and $c$ is the speed of light in vacuum. For a master laser operating at 200-MHz repetition rate, if we have a deviation of only $\Delta f_{rep}$ = 1 kHz and a length difference of $\Delta d$ =10 cm, we will observe a timing error of ~1.7 fs between the two slave lasers. Therefore, one has to pay considerable attention to the long-term stability of the master laser and to the length of the uncompensated beam paths.

Furthermore, synchronization architecture and simplicity also play an important role in order to minimize the number of noise sources present in the system. The "continuous-wave" (cw) approach widely used in optical clock comparison and remote laser synchronization[9,24] requires additional monochromatic ultra-stable transfer lasers whose optical frequency must be within the bandwidths of the slave mode-locked lasers to get the heterodyne beat. Therefore, many transfer lasers at different-wavelengths are employed[24] to achieve a multi-color synchronous mode-locked laser network which increases the complexity of the feedback system and the number of noise sources. Furthermore, this also makes the fiber transmission over long distances impractical for wavelengths other than 1550 nm if lasers operating at different wavelengths have to be synchronized. Here, we demonstrate a synchronous multi-color mode-locked laser network distributed over 4.7-km distance with daily sub-femtosecond timing drift. The laser network is based on the pulsed timing synchronization scheme[27-29] which uses a mode-locked laser with 1550-nm center wavelength as its master oscillator. Unlike the cw approach, the optical signal of the master laser is transmitted directly to the slave lasers via timing-stabilized fiber links where synchronization has been achieved by four-wave mixing between the link outputs and the slave lasers. Measured out-of-loop timing drift between two remote slave lasers is only 0.6 fs RMS over 40 h corresponding to a relative timing instability of $1.2\times10^{-20}$ at 70000-s averaging time. To the best of our knowledge, this is the lowest timing drift measured over multiple days between two slave lasers synchronized across a fiber link network. Moreover, we measure the short term stability of the system above 1 Hz and show that the total integrated jitter between the slave lasers is only 1.3 fs RMS integrated between 1 Hz and 1 MHz. Our noise analysis reveals that the slave laser's inherent jitter is the most dominant noise source with 1.26-fs RMS. This suggests that the precision of the laser network is not limited by our synchronization scheme, and it could be improved even further with the use of lower noise mode-locked lasers. In this paper, we first describe the experimental implementation of the synchronous mode-lock laser network and show the measurement results. Then, we present our timing jitter analysis to uncover the limiting noise factors within the system.

**Experimental setup**

We have realized a synchronous mode-locked laser network experimentally as described in Fig. 2(a). Three distant lasers are synchronized in a star network topology across two different timing-stabilized fiber links with 4.7-km total length. We apply the pulsed timing synchronization scheme[27-29] using balanced optical cross correlators (BOC)[19] to stabilize the signal delay through polarization-maintaining (PM) fiber links as well as to synchronize the remote slave lasers at the link outputs. Our master laser is a commercially available mode-locked laser (Origami-15 from OneFive GmbH) operating at 216.67-MHz repetition rate with 1554-nm center wavelength and 172-fs pulse duration. It has a free-running timing jitter of only 0.4-fs RMS integrated between 1 kHz and 1 MHz[30]. To avoid long-term frequency drifts, the repetition rate of the master laser is locked to a radio-frequency (RF) synthesizer with lower phase noise at small offset frequencies (below 100 Hz) using a phase-locked loop (PLL in Fig. 2(b)). The output of the master laser is split into two independent timing links with a total length of 4.7 km. Each link consists of a PM dispersion-compensated fiber spool, a PM fiber stretcher, and a motorized delay stage (see "Timing link" schematic in Fig. 2(b)).

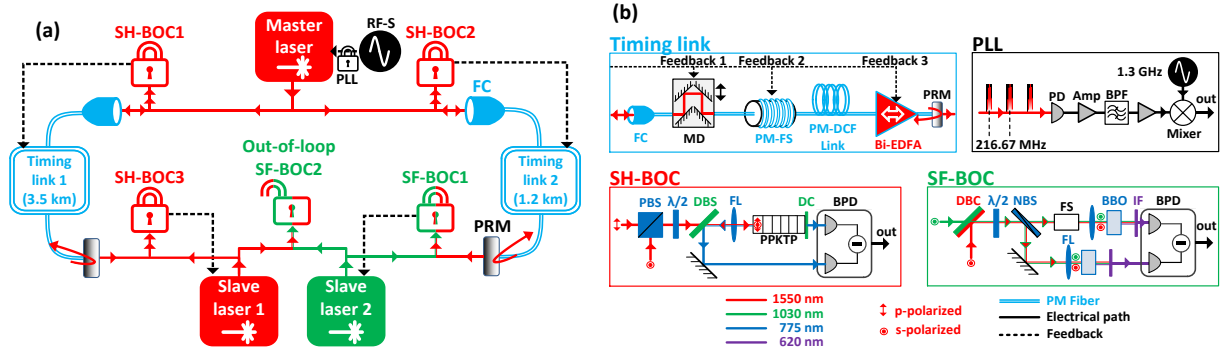

**Figure 2.** (a) Experimental setup of the synchronous mode-locked laser network. All BOCs are symbolized with "lock symbols" shown under their abbreviations. A sealed lock refers to an in-loop detector; whereas an open lock corresponds to an out-of-loop detector. (b) Detailed schematics of the individual elements. Abbreviations: SH-BOC: second-harmonic balanced optical cross-correlator; SF-BOC: sum-frequency BOC; PLL: phase-locked loop; RF-S: RF synthesizer; FC: fiber collimator; PRM: partially reflecting mirror; PBS: polarization beam splitter; λ/2: half-wave plate; DBS: dichroic beam splitter; FL: focus lens; PPKTP: periodically poled potassium titanyl phosphate crystal; DC: dichroic coating; BPD: balanced photodetector; PD: photodetector; AMP: voltage amplifier; BPF: 1.3-GHz electronic bandpass filter; DBC: dichroic beam combiner; NBS: neutral 50:50 beam splitter; FS: fused-silica plate; BBO: barium borate crystal; IF: interference filter; MD: motorized delay; PM-FS: polarization-maintaining fiber stretcher; PM-DCF: PM dispersion-compensated fiber; Bi-EDFA: PM bi-directional erbium doped fiber amplifier.

A partially reflecting mirror at the end of each link reflects 10% of the optical power back to the link input. A bi-directional Erbium doped fiber amplifier before the mirror amplifies the pulse energy for forward and backward propagation to provide sufficient power for link-stabilization at the link input and for laser synchronization at the link output. Upon roundtrip propagation, the timing of the pulses is compared against new pulses from the master laser using type-II second-harmonic (SH) BOCs designed for 1554-nm wavelength (SH-BOC in Fig. 2(b)). Optical cross-correlation is achieved through the birefringence of the nonlinear crystal between the two orthogonally polarized input pulses. In this way, SH-BOC1 and SH-BOC2 detect propagation delay fluctuations in the timing links and generate voltage responses which are fed back to the motorized delays and fiber stretchers to compensate for long-term timing drift and fast timing jitter, respectively (Feedbacks 1 and 2 in Fig. 2(b)). Furthermore, a third feedback is sent to the pump current of the fiber amplifiers to eliminate the link power fluctuations caused by the beam misalignments upon the movement of the motorized delays. As detailed in our previous work[23], optical power variations induce timing jitter and drift at the fiber link output through a composite effect of residual dispersion and fiber nonlinearities which cannot be corrected by the BOC detection scheme. This feedback mechanism ensures attosecond-precision timing link stabilization by minimizing the noise contribution of these effects.

The outputs of the timing links are used for the remote synchronization of two slave mode-locked lasers. Slave laser 1 (Origami-15 from OneFive GmbH) is identical to the master laser. SH-BOC3 is built to synchronize slave laser 1 with the output of timing link 1 by tuning the repetition rate via its intracavity piezoelectric (PZT) actuated mirror. Slave laser 2 (Origami-10 from OneFive GmbH), on the other hand, operates at 54.17-MHz repetition rate with an output optical spectrum centered at 1030 nm which is widely used for the seeding of Yb-based amplifier chains, such as the pump-probe laser system at the European XFEL[31]. To synchronize slave laser 2 with timing link 2 output, a type-I sum-frequency BOC (SF-BOC1) is built for 1030-nm and 1554-nm operation using two BBO crystals (see SF-BOC schematic in Fig. 2(b)). Time delay difference between the two arms of the SF-BOC is adjusted with a glass plate to maximize the timing sensitivity of the balanced detection. A fraction of the output power from each slave laser is spared for an out-of-loop timing detector (i.e., SF-BOC2 in Fig. 2(a)) to evaluate the timing precision of the laser network.

The experimental setup consumes a total space of two optical tables. The master laser, SH-BOC1 and SH-BOC2 are placed on the first optical table which is actively temperature stabilized and covered with an acoustically isolated enclosure. The fiber links are placed outside of the enclosure and exposed to the environmental fluctuations in the laboratory. The link outputs connect to the second optical table where the slave lasers, SH-BOC3, SF-BOC1 and SF-BOC2 are constructed inside another optical enclosure.

## Experimental results

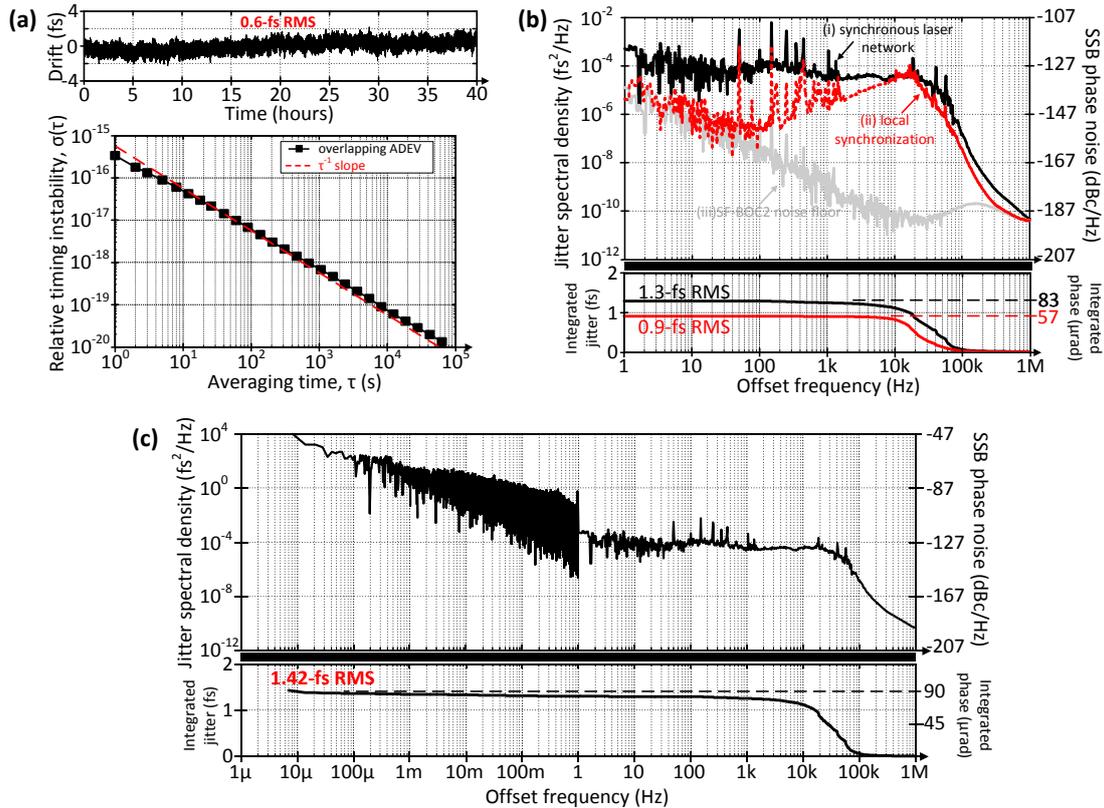

**Figure 3.** Experimental results of the synchronous mode-locked laser network. (a) Long-term performance; top graph shows the relative timing drift data sampled at 2 Hz and underneath is the relative timing instability vs. averaging time. Please note, that the measurement noise below 1 Hz is only ~0.8 as RMS (see Methods section). (b) Short-term performance above 1 Hz; relative timing jitter spectral density and integrated timing jitter of (i) the laser network; (ii) the local synchronization between the slave lasers and (iii) the noise floor of SF-BOC2. Please note that detector noise floor is only 12 as RMS integrated for 1 Hz – 1 MHz. (c) Timing jitter and phase noise for the complete measurement range from 7 μHz up to 1 MHz. Right axes of (b) and (c) indicate the equivalent single-sideband (SSB) phase noise rescaled to 10-GHz carrier frequency.

Figure 3 shows the experimental results measured with the out-of-loop timing detector, which performs the intensity cross-correlation between the output pulse trains of the two remotely synchronized slave lasers. The first set of results in Fig. 3(a) describes the long-term performance of the synchronous mode-locked laser network. Relative timing drift between the slave lasers is only 0.6 fs RMS over 40 h of continuous measurement with no apparent long-term fluctuations. This corresponds to the lowest timing instability ever measured between two mode-locked lasers remotely synchronized to a common reference over a multi-kilometer distance. The second graph in Fig. 3(a) shows the overlapping Allan deviation calculated from the timing drift data. The laser network exhibits a relative timing instability of 5×10$^{-17}$ only after 10-s averaging time ($\tau$) and crosses the 20$^{th}$ decimal instability border after ~8000 s by following a slope very close to $\tau^{-1}$. This slope characteristic indicates that the laser network does not have excess timing drift even for long measurement durations.

Timing jitter analysis on short time scales is essential to examine the noise contributions of the system elements for further improvement. We measure the timing jitter spectral density of the synchronous laser network with a baseband analyzer which decomposes the voltage output of SF-BOC2 into its frequency components. The black curve in Fig. 3(b) shows the measured out-of-loop timing jitter. If we exclude the power line spurs observed at the harmonics of 50 Hz, the complete spectrum is below a jitter spectral density of 8×10$^{-4}$ fs$^2$/Hz. This corresponds to a relative phase noise of only -118 dBc/Hz at 1-Hz between two 10-GHz oscillators located 4.7-km far from each other. Furthermore, total timing jitter of the laser network is only 1.3-fs (±0.006 fs) integrated from 1 Hz to 1 MHz. This corresponds to a relative phase error of 83 µrad (±0.4 µrad) at 10-GHz carrier frequency where the uncertainty in the measured values arises from the measurement error of the timing sensitivity given in the Methods section. Please note that the noise above 1 MHz is obscured by the detector noise floor and negligible. This can be clearly seen from the integrated timing jitter of the laser network in Fig. 3(b) which has the most dominant noise contribution from the offset frequencies between 10 kHz and 100 kHz.

To investigate the high frequency noise limitations of the network, the slave lasers are disengaged from their links and locally synchronized to each other using SF-BOC2. The red curve in Fig. 3(b) shows the in-loop jitter spectral density of the local synchronization which closely follows the network results for frequencies between 10 kHz and 1 MHz. As the in-loop detector obscures the noise between the slave lasers for frequencies lower than the locking bandwidth (~10 kHz), timing jitter spectral density of the local synchronization shows larger deviation from the network results in this region (see the red dotted curve in Fig. 3(b) below 10 kHz). Nevertheless, integrated timing jitter of the local synchronization is as high as 0.9 fs RMS pointing out the inherent timing jitter of the slave lasers beyond the locking bandwidth as the prominent noise contribution.

Figure 3(c) shows the timing jitter spectrum of the laser network for the complete measurement range from 7 µHz up to 1 MHz (i.e., from 1 µs to 40 h in time domain). The network has a total timing jitter of 1.42 fs RMS. If we exclude the noise contribution above 10 kHz, the total timing jitter stays in the attosecond regime; i.e., 670 as RMS integrated from 1 kHz down to 7 µHz.

**Sources of timing jitter**

In this section, we present a timing jitter analysis of the synchronous mode-locked laser network, in order to comprehend the limitations of timing transfer using optical pulse trains and fiber links. A straightforward approach would be to measure and quantify the timing jitter generated by the noise sources present in the system. For instance, timing jitter measurement of the local synchronization shown in Fig. 3(b) indicates the inherent noise of the slave lasers. However, this does not determine their actual jitter contribution when used in a complex system such as our synchronous mode-locked laser network where many feedback loops are employed. An effective way to identify sources of timing jitter arising from such complex interdependencies is to investigate the feedback flow between setup elements[32]. Here, we determine the noise sources of our laser network with a comprehensive feedback loop analysis adapted from our recent work[32]. The detailed derivation of the noise sources is given in the Supplementary Information. According to our analysis, the out-of-loop timing jitter spectral density measured between the slave lasers is given as:

$$\overline{J_{out}^2} = |C_M|^2\overline{J_M^2} + |C_{S1}|^2\overline{J_{S1}^2} + |C_{S2}|^2\overline{J_{S2}^2} + |C_{E,L1}|^2\overline{J_{E,L1}^2} + |C_{E,L2}|^2\overline{J_{E,L2}^2} + |C_{N,S1}|^2\overline{J_{N,S1}^2} \quad (2)$$
$$+ |C_{N,S2}|^2\overline{J_{N,S2}^2} + |C_{N,L1}|^2\overline{J_{N,L1}^2} + |C_{N,L2}|^2\overline{J_{N,L2}^2}$$

In total, there are nine uncorrelated noise sources in the out-of-loop timing jitter: $J_M$, $J_{S1}$ and $J_{S2}$ are the free-running timing jitter of the master laser, slave laser 1 and slave laser 2 respectively. $J_{E,L1}$ and $J_{E,L2}$ are the integrated environmental noise imposed on timing link 1 and timing link 2 for one-way transmission. $J_{N,S1}$ and $J_{N,S2}$ represent the electronic noise of the feedback controls based on SH-BOC3 and SF-BOC1 synchronizing the slave lasers with the timing link outputs; whereas $J_{N,L1}$ and $J_{N,L2}$ are the electronic noise of the feedback controls based on SH-BOC1 and SH-BOC2 stabilizing the timing links. The terms in front of each noise source ($C_M$, $C_{S1}$, $C_{S2}$, $C_{E,L1}$, $C_{E,L2}$, $C_{N,S1}$, $C_{N,S2}$, $C_{N,L1}$, and $C_{N,L2}$) are frequency-dependent complex "jitter transfer functions" which determine the contribution of the noise sources to the out-of-loop timing jitter of the laser network. Please note that, all terms in Eq. (2) can be either measured or calculated using experimental parameters (refer to the Methods section).

Fig. 4(a) shows the results of the timing jitter analysis from 100 Hz to 100 kHz and the annotation on the right provides the color code of the jitter spectral densities and their integrated jitter below. Note that the lower limit for the offset frequency is set by the free-running timing jitter of the master laser which could not be resolved below 100 Hz, whereas the higher frequency limit is set by the environmental noise of the timing links which is smaller than the measurement noise floor above 100 kHz. As we have suspected from the local synchronization results shown in Fig. 3(b), the most prominent noise contribution originates from the slave laser 2 and it dominates the complete timing jitter spectrum with 1.26-fs RMS integrated jitter. The environmental noise on the timing links are well suppressed by the feedback loops for offset frequencies below 10 kHz. However, one has to pay attention to the PZT resonances which makes the environmental jitter of timing link 1 the second largest noise source with 0.28-fs RMS (see the jitter peak around 70-kHz of the dark green curve in Fig. 4(a)). The master laser and slave laser 1 have minor contributions with 0.11-fs RMS and 0.10-fs RMS integrated jitter due to their lower inherent noise when compared to slave laser 2. Furthermore, we can see the benefit of low noise floors provided by BOC-locking schemes since the least dominant jitter contributions in our system are the four electronic noise terms.

Finally, Fig. 4(b) compares the sum of all noise sources in Eq. (2) with the measurement results. The calculated jitter spectrum is in a very good agreement with the experiment and reveals many of the measured jitter structures such as the servo bump around 20 kHz and electronic noise spurs below 1 kHz. Integration of the spectrum results in 1.31-fs total jitter which also agrees quite well with the experiment. Nonetheless, there are some small deviations between the measurement and the calculation especially for offset frequencies between 15 kHz and 40 kHz. This frequency range is highly influenced by the PZT resonances in the complete feedback system. We believe that the agreement between the calculation and the measurement could be improved even further by including those resonance frequencies in the modeling.

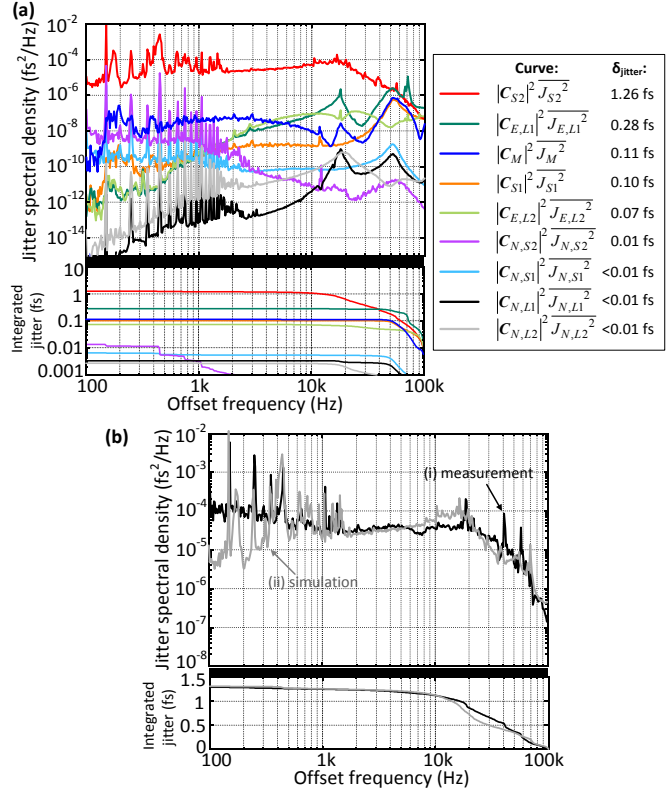

**Figure 4.** Noise sources of the synchronous laser network. (a) Jitter spectral densities of each noise source and their integrated timing jitter in logarithmic scale. The annotation on the right shows the color code of the graphs. (b) (i) measured out-of-loop timing jitter, (ii) sum of all calculated noise sources given in (a).

## Conclusion

In conclusion, we report a synchronous, multi-color laser network over 4.7-km distance with long-term sub-femtosecond stability. We detect a record-low out-of-loop timing drift of 0.6 fs RMS between two remote lasers over 40 h. Allan deviation calculation of the drift data reveals a relative timing instability of $5\times10^{-17}$ in 10 s reaching $1.2\times10^{-20}$ in 70000 s by following a deterministic $\tau^{-1}$ slope. This is a clear indication that the synchronous laser network does not accumulate excess timing drift and hence the timing instability between two remote slave lasers decreases inversely with the increasing integration time.

We also measure the out-of-loop jitter spectral density of the network which has only 1.3-fs RMS integrated jitter from 1 Hz up to 1 MHz. By pursuing a comprehensive feedback loop analysis, we reveal nine independent noise sources within the laser network. The inherent jitter of the slave laser 2 is the most dominant source with 1.26-fs RMS total jitter, whereas the electronic noise arising from the feedback loops contributes only a total jitter of 0.01 fs RMS due the high signal to noise ratio of BOC-locking systems. From the detailed analysis, we conclude that the high frequency performance of the synchronous mode-locked laser network can be improved even further to the attosecond regime by developing lasers with lower inherent noise.

We believe that such a synchronous laser network can provide X-ray laser research facilities with the highest temporal resolution and lead to new discoveries in ultrafast molecular dynamics. Besides, the demonstrated network can also be implemented in other ambitious large-scale scientific explorations, such as comparison of distant optical clocks[12], sensitive imaging of low temperature black bodies using multi telescope arrays[33], and geodetic modelling of the Earth surface and atmosphere using very-long-baseline interferometers[34].

## Methods

**Out-of-loop timing measurements.** Once all timing detectors and their feedback loops in the experimental setup (see PLL, SH-BOC1-3 and SF-BOC1 in Fig. 2(a)) are activated, the repetition rates of the slave lasers are locked to the master laser remotely via the timing links. Then, we use a motorized delay stage to temporally overlap the two optical pulse trains and measure their relative timing with the free-running SF-BOC2. In this case, the output of SF-BOC2 is a DC voltage. To measure the timing sensitivity, we sweep the relative delay between the pulse trains, while the response voltage of SF-BOC2 is recorded. The linear slope around the zero-crossing of the response voltage is the timing sensitivity with the units of mV/fs. Five independent measurements of SF-BOC2's timing sensitivity result in a mean value of 14.96 mV/fs with a measurement error of ±0.066 mV/fs due to slope uncertainty. Long-term drift data in Fig. 3(a) are measured by filtering the out-of-loop signal with a 1-Hz low-pass anti-aliasing filter and recording it at 2-Hz sampling rate. The noise floor of this measurement is governed by the noise of the measurement device (NI USB-6211) which is specified as 12 μV RMS. Considering the timing sensitivity of the out-of-loop BOC, the measurement noise below 1 Hz is only ~0.8 as RMS. The jitter spectral density data in Fig. 3(b) are the baseband power spectrum of the output voltage of SF-BOC2 measured by a signal source analyzer (Agilent E5052B). In Fig. 3(c), the data above 1 Hz are identical to the data set (i) in Fig. 3(b); whereas, the data below 1 Hz are the Fourier frequency components of the drift data in Fig. 3(a).

**Noise sources and their jitter transfer functions.** Analytical derivation of Eq. (2) is given in the Supplementary Information. Since the master laser and slave laser 1 are identical, we use the experimental data set from our former publication[30] for both lasers (i.e., $\overline{J_M^2} = \overline{J_{S1}^2}$). Inherent timing jitter of slave laser 2 ($\overline{J_{S2}^2}$) is derived from the local synchronization data shown in Fig. 3(b) and detailed in the Supplementary Information. The environmental noise sources imposed on the links ($\overline{J_{E,L1}^2}$, $\overline{J_{E,L2}^2}$) are obtained experimentally by measuring the free-running outputs of SH-BOC1 and SH-BOC2 with the signal source analyzer when their feedback loops are disengaged. Then, the electronic noise terms ($\overline{J_{N,S1}^2}, \overline{J_{N,S2}^2}, \overline{J_{N,L1}^2}, \overline{J_{N,L2}^2}$) and the jitter transfer functions ($|C_M|^2, |C_{S1}|^2, |C_{S2}|^2, |C_{E,L1}|^2, |C_{E,L2}|^2, |C_{N,S1}|^2, |C_{N,S2}|^2, |C_{N,L1}|^2, |C_{NL,2}|^2$) are calculated using the experimental parameters as described in the Supplementary Information. Fig. S3 and Fig. S4 in the Supplementary Information show the jitter spectral densities of all noise sources and their calculated jitter transfer functions.

**Data availability.** The authors declare that all the data in this manuscript are available.

**Acknowledgments**
The authors thank Dr. Laurens Wissmann, Dr. Jinxiong Wang and Dr. Maximilian Lederer for allowing the use of the laser Origami-10 in the experiment. This work has been supported by the European Research Council under the European Union's Seventh Framework Programme (FP/2007-2013) / ERC Grant Agreement n. 609920 and the excellence cluster "the Hamburg Centre for Ultrafast Imaging" of Deutsche Forschungsgemeinschaft.

## Author Contributions

F.X.K. initiated the project. K.Ş., M.X., M.Y.P. and F.X.K. conceived and designed the experiments. K.Ş. and M.X. performed the experiments and investigated sources of timing jitter in the system. All authors contributed to the preparation of the manuscript.

## Additional Information

**Supplementary information:** accompanies the paper on the website of the journal.
**Competing interests:** The authors declare no competing interests.